\documentclass[12pt,preprint]{aastex}
\usepackage[english]{babel}
\usepackage{amsmath}
\usepackage{amssymb}
\usepackage{graphicx}
\usepackage[dvipdf]{color}

\newcommand{\be}{\begin{equation}}
\newcommand{\ee}{\end{equation}}
\newcommand{\beq}{\begin{eqnarray}}
\newcommand{\eeq}{\end{eqnarray}}

\begin{document}

\title{Investigating superconductivity in neutron star interiors with glitch models}
\author{B.~Haskell} 
\affil{Astronomical Institute ``Anton Pannekoek'', University of
  Amsterdam, Postbus 94249, 1090 GE Amsterdam, The Netherlands\\
Max-Planck-Institut f\"{u}r Gravitationsphysik, Albert-Einstein-Institut, Am M\"{u}hlenberg 1, 14476 Golm, Germany}
\author{P.~M.~Pizzochero and S.~Seveso}
\affil{Dipartimento di Fisica, Universit\'{a} degli Studi di Milano, Via Celoria 16, 20133 Milano, Italy\\
Istituto Nazionale di Fisica Nucleare, sezione di Milano, Via Celoria 16, 20133 Milano, Italy}

\begin{abstract}
\noindent

The high density interior of a neutron star is expected to contain superconducting protons and superfluid neutrons. Theoretical estimates suggest that the protons will form a type II superconductor in which the stellar magnetic field is carried by flux tubes. The strong interaction between the flux tubes and the neutron rotational vortices could lead to strong 'pinning', i.e. vortex motion could be impeded. This has important implications especially for pulsar glitch models as it would lead to a large part of the vorticity of the star being decoupled from the 'normal' component, to which the electromagnetic emission is locked. In this paper we explore the consequences of strong pinning in the core on the 'snowplow' model for pulsar glitches (Pizzochero 2011), making use of realistic equations of state and relativistic background models for the neutron star. We find that in general a large fraction of pinned vorticity in the core is not compatible with observations of giant glitches in the Vela pulsar. The conclusion is thus that either most of the core is in a type I superconducting state or that the interaction between vortices and flux tubes is weaker than previously assumed. 

\end{abstract}
\maketitle

\section{Introduction}

Neutron Stars (NSs) allow us to probe the state of matter in some of the most extreme conditions in the universe. Not only can the density in the interior of these very compact objects exceed nuclear saturation density, but NSs also host some of the strongest magnetic fields in nature, with intensities of up to $\approx 10^{15}$ G for magnetars.
Not surprisingly modelling such complex objects requires the use of some poorly understood physics. 

In particular the star will rapidly cool below the critical temperature for the neutrons to be superfluid and the protons to be superconducting. The protons of the outer core are predicted to form a type II superconductor \citep{Migdal, Baym}, in which the magnetic flux is confined to flux tubes, inside which the magnetic field strength is of the order of the lower critical field for superconductivity, $B_c \approx 10^{15}$ G. However above a critical density of approximately $\rho_c\approx 3\times 10^{14}$ g/cm$^3$ one expects a transition to type I superconductivity, in which the formation of fluxtubes is no longer favourable but rather the magnetic field is contained in regions of normal protons \citep{Sedrakian05}. Given that the critical density for this transition is easily reached in NS interiors it is possible that a sizeable portion of the star may in fact be in a type I superconducting state \citep{Jones06}.

The dynamics of the outer core plays a crucial role in the interpretation of various astrophysical phenomena, such as pulsar glitches, timing noise, precession and fluid oscillations. In particular pulsar glitches are sudden increases in the otherwise steadily decreasing rotational frequency of a pulsar. Although their origin is still debated it is generally thought that these phenomena are the direct manifestation of a superfluid component inside the star, which is only weakly coupled to the normal component due to the interaction between the quantized neutron vortex lines and the charged particles in the crust or core. In particular if vortices can ``pin'' (i.e. are strongly attracted) to the Coulomb lattice in the crust, they can decouple the neutron superfluid from the normal component (to which the electromagnetic emission is locked) and their sudden depinning will give rise to a rapid transfer of angular momentum, i.e. a glitch \citep{AndItoh75, Alpar77, Pines80, Alpar81, Anderson82}. Recent work has shown that this scenario can successfully account for the distribution in glitch sizes and waiting times \citep{Melatos1,Melatos2,Melatos3,Andrewnew} and describe the size and relaxation timescales of giant glitches in the Vela pulsar \citep{Piz11,Hask11}.

An important issue to address is, however, whether vortices will only pin to the crustal lattice or whether they are pinned to flux tubes if the outer core is in a type II superconducting state \citep{Link03}, thus effectively decoupling a large fraction of the stellar moment of inertia from the crust. Furthermore if vortices are pinned in the core this is likely to lead to the onset of turbulence and may play an important role in pulsar 'timing noise' \citep{Link11}. The interaction between flux tubes and vortices can also have a strong impact on the gravitational wave driven r-mode instability \citep{Ho11,Hask12} and on NS precession \citep{Link03}.

In this Letter we investigate the effect of vortex pinning in the core on the ``snowplow'' glitch model of \citet{Piz11}. We extend the model to realistic equations of state and relativistic stellar models, as in \citet{Stefano2} and show that, in general, one cannot fit the size and postglitch jumps in frequency derivative of Vela giant glitches if a large portion of the core vortices are pinned.
This points towards the fact that most of the core could in fact be in a type I superconducting state, or that the vortex/flux tube interaction is weaker than previously assumed, as some microphysical estimates suggest \citep{Babaev}.

Furthermore \citet{Glamp11} recently showed that vortex pinning in the core is likely to be a short lived phenomenon that may only be relevant in a short period of a NS's life and in magnetars, and in their hydrodynamical model of giant pulsar glitches \citet{Hask11} also find that vortex pinning in the core is inconsistent with the observed post-glitch relaxation timescales in the Vela.

\section{The ``snowplow'' model}
\label{pinc}

The starting point of our investigation will be the ``snowplow'' model for glitches of \citet{Piz11}, which we briefly review here. We take the NS to be a two component system, where one of the components, the so-called ``normal'' component, is given by the crust and all charged components tightly coupled to it by the magnetic field. The other, the ``superfluid'', is given by the superfluid neutrons in the core and crust. The superfluid rotates by forming an array of quantized vortices which carry the circulation and mediate an interaction between the two components known as Mutual Friction, which in the core can couple the two fluids on timescales of seconds \citep{AndSid}.
Vortices can, however, also be {\it pinned} to ions in the crust or flux-tubes in the core \citep{AndItoh75, Alpar77, Pines80, Alpar81, Anderson82, Rud, Link03}. As a consequence vortex motion is impeded and the superfluid cannot spin-down, effectively decoupling it from the normal component which is spinning down due to electromagnetic emission. 
If a lag builds up between the superfluid and the normal component this will, however, give rise to a Magnus force acting on the vortices, which takes the form 
$\mathbf{f}_m=\kappa \rho_s{\boldsymbol{\hat{\Omega}}}\times (\mathbf{v}_v-\mathbf{v}_s)$, where $\mathbf{f}_m$ is the force {\it per unit length}, $\kappa=1.99\times 10^{-3}$ cm$^2$s$^{-1}$ is the quantum of circulation, $\rho_s$ is the superfluid density, ${\boldsymbol{\hat{\Omega}}}$ is the unit vector pointing along the rotation axis, $\mathbf{v}_v$ is the velocity of the vortex lines and $\mathbf{v}_s$ is the velocity of the superfluid. We assume that the neutrons are superfluid throughout the star and take $\rho_s=(1-x_p)\rho$, with $x_p$ the proton fraction calculated by \citet{Zuo}. Once the Magnus force integrated over a vortex exceeds the pinning force, the vortex will unpin and be free to move out.

We follow the procedure of \citet{Stefano2} and integrate the relativistic equations of stellar structure for two realistic equations of state, SLy \citep{DH01} and GM1 \citep{GM1}. We assume straight vortices that cross through the core \citep{Zhou} and for the pinning force per unit length $\mathbf{f}_p$ we use the realistic estimates of \citet{Gpaper,Gtesi}. Balancing the pinning force to the Magnus force and integrating over the vortex length allows us to calculate the lag at which the vortices will unpin in different regions. The normalization of $\mathbf{f}_p$ is chosen in such a way as to give an inter-glitch waiting time $T_g=\Delta\Omega_\mathrm{max}/|\dot{\Omega}|$ of approximately 2.8 yrs for the Vela pulsar, where $\Delta\Omega_\mathrm{max}$ is the maximum of the critical unpinning lag.

If there is no pinning in the core vortices will unpin and move out toward the crust, where they encounter a steeply increasing pinning potential and repin.  This leads to the creation of a thin vortex sheet that moves towards the peak of the potential, the so-called ``snowplow'' effect. Once the maximum of the critical lag has been reached the vortices can no longer be held in place and the excess vorticity is released catastrophically, exchanging angular momentum with the normal component and giving rise to a glitch \citep{Piz11}.  We assume that this is the mechanism that gives rise to giant glitches, i.e. glitches with steps in the spin rate $\Delta\Omega_{gl}\approx 10^{-4}$ rad that are observed in the Vela and other pulsars \citep{Espinoza}. Smaller glitches are likely to be triggered by crust quakes or random vortex avalanches \citep{Melatos1, Melatos2, Melatos3}.

We can easily calculate the number of vortices in the vortex sheet once it has reached the peak of the potential as $N_v=\frac{2\pi}{\kappa}r^2_{\mathrm{max}}\Delta\Omega_{\mathrm{max}}$, where $r_{\mathrm{max}}$ is the cylindrical radius at which the maximum of the critical lag is located and $\Delta\Omega_{\mathrm{max}}$ the value of said maximum. The angular momentum exchanged as the vortices move out and annihilate is then given by:
\be
\Delta L_{\mathrm{gl}}=2\kappa Q N_v\int_{r_{\mathrm{max}}}^{R_c} x dx\int_0^{l(x)/2} \rho(\sqrt{x^2 + z^2}) dz\label{moment}
\ee
where $Q=I_s/I_{tot}$ is the superfluid fraction of the moment of inertia, $R_c$ is the radius of the inner crust where the vortices annihilate (taken at neutron drip density), $l(x)$ is the length of a vortex at a given cylindrical radius $x$ and $\rho$ is the density. 
The glitch observables can then be derived as $\Delta\Omega_{gl}={\Delta L}/{I_{\mathrm{tot}}[1-Q(1-Y_{gl})]}$ and ${\Delta\dot{\Omega}_{gl}}/{\dot{\Omega}_{\infty}}={[Q(1-Y_{gl})]}/{[1-Q(1-Y_{gl})]}$, where $\Delta\Omega_{gl}$ is the step in angular velocity due to the glitch and ${\Delta\dot{\Omega}_{gl}}/{\dot{\Omega}_{\infty}}$ is the {\it instantaneous} step in the spin-down rate immediately after the glitch, relative to the steady state pre-glitch spindown rate $\dot{\Omega}_\infty$. We have also introduced the parameter $Y_{gl}$ which represents the fraction of superfluid moment of inertia which is coupled to the crust during the glitch. Given that the rise time $\tau_r$ is very short (less than a minute \citep{Dod}) it is likely that only a small fraction of the core will be coupled to the crust on this short timescale, with the rest of the star recoupling gradually on longer timescales and giving rise to the observed exponential post-glitch relaxation (see \citet{Hask11} for a detailed discussion of this issue). The best observational upper limits on the rise time are $\tau_r < 40$ s \citep{Dod} from the Vela 2000 glitch, while an interesting lower limit of $\tau > 10^{-4}$ ms can be derived from the non detection of a GW signal from the Vela 2006 glitch \citep{Andrewnew}. Theoretical estimates, on the other hand, give $\tau_r\approx 1-10$ s \citep{Hask11}, which thus easily allow for the angular momentum in (\ref{moment}) to be exchanged during the short rise times observed in radio pulsars.


Let us now consider the motion of a vortex if the NS core is a type II superconductor. As a vortex approaches a flux tube its magnetic energy will increase if they are alligned or decrease if they are antialigned reasulting in an energy per intersection of approximately $E_p\approx 5$ MeV  \citep{Rud}. Note that we have neglected the contribution associated with the reduction of the condensation energy cost if a vortex and a flux tube overlap. This leads to an energy cost per interesection slightly smaller than that estimated above (of the order $E_p\approx 0.1 -1 $ MeV) \citep{Rud, TrevAlp}. Vortex motion is thus impeded by the flux tubes, that provide an effective pinning barrier unless the vortices have enough energy to cut through them.
The corresponing pinning force per unit length of a vortex has been estimated to be $f_p\approx 3\times 10^{15} B_{12}^{1/2}$ dyn cm$^{-1}$ \citep{Link03}, and is balanced by the Magnus force for a critical relative velocity of $w_c\approx 5\times 10^3 B_{12}^{1/2}$ cm s$^{-1}$. This leads to a critical lag (at a radius of 10 km) $\Delta\Omega_c\approx 5\times 10^{-3} B_{12}^{1/2}$ rad, where we have assumed an average density for the core of $\rho=3\times 10^{14}$ g cm$^{-3}$.
Given the large value of the critical lag, comparable to what could be built up in-between Vela glitches, a substantial part of the vorticity in the core could be pinned. 

To account for this effect we shall assume that a fraction of the vorticity in the core is pinned and does not contribute to the angular momentum stored in the vortex sheet. This is equivalent to assuming that all the vorticity within a radius $R_v=\eta R_b$ is frozen, with $R_b$ the radius of the base of the crust and $\eta$ a free parameter. We thus define a fraction of pinned vorticity in the core as $\xi=(R_v^2)/(r_{\mathrm{max}}^2)$. The total number of vortices in the vortex sheet before the glitch scales accordingly:
\be
N_v=(1-\xi)\frac{2\pi}{\kappa}(r^2_{\mathrm{max}})\Delta\Omega_{\mathrm{max}}\label{vortex}
\ee
By using (\ref{vortex}) in (\ref{moment}) we can obtain the angular momentum exchanged during the glitch and by fitting the size of a glitch, $\Delta\Omega_{\mathrm{gl}}$, we can derive the coupled fraction of superfluid
\be
Y_{gl}=\frac{1}{Q(1-\xi)}\left[\frac{\Delta L}{\Delta\Omega_{\mathrm{gl}}I_{\mathrm{tot}}}+Q-1\right].\label{step1}
\ee
The instantaneous step in the frequency derivative then follows from
\be
\frac{\Delta\dot{\Omega}_{gl}}{\dot{\Omega}_{\infty}}=\frac{Q(1-\xi)(1-Y_{gl})}{1-Q[1-(1-\xi) Y_{gl}]}\label{stepdot}
\ee

\begin{figure*}
\centerline{\includegraphics[scale=0.5]{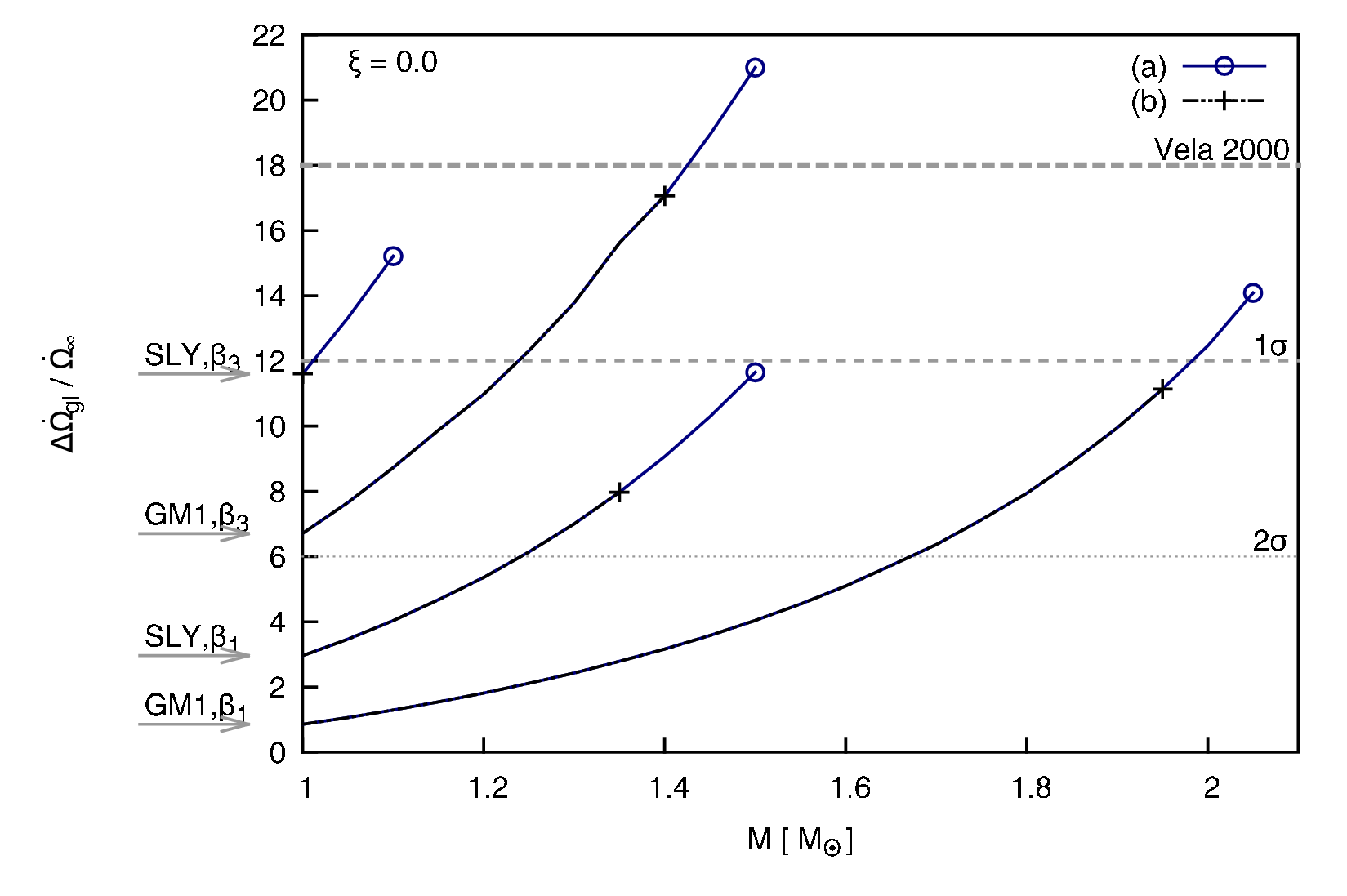}\includegraphics[scale=0.5]{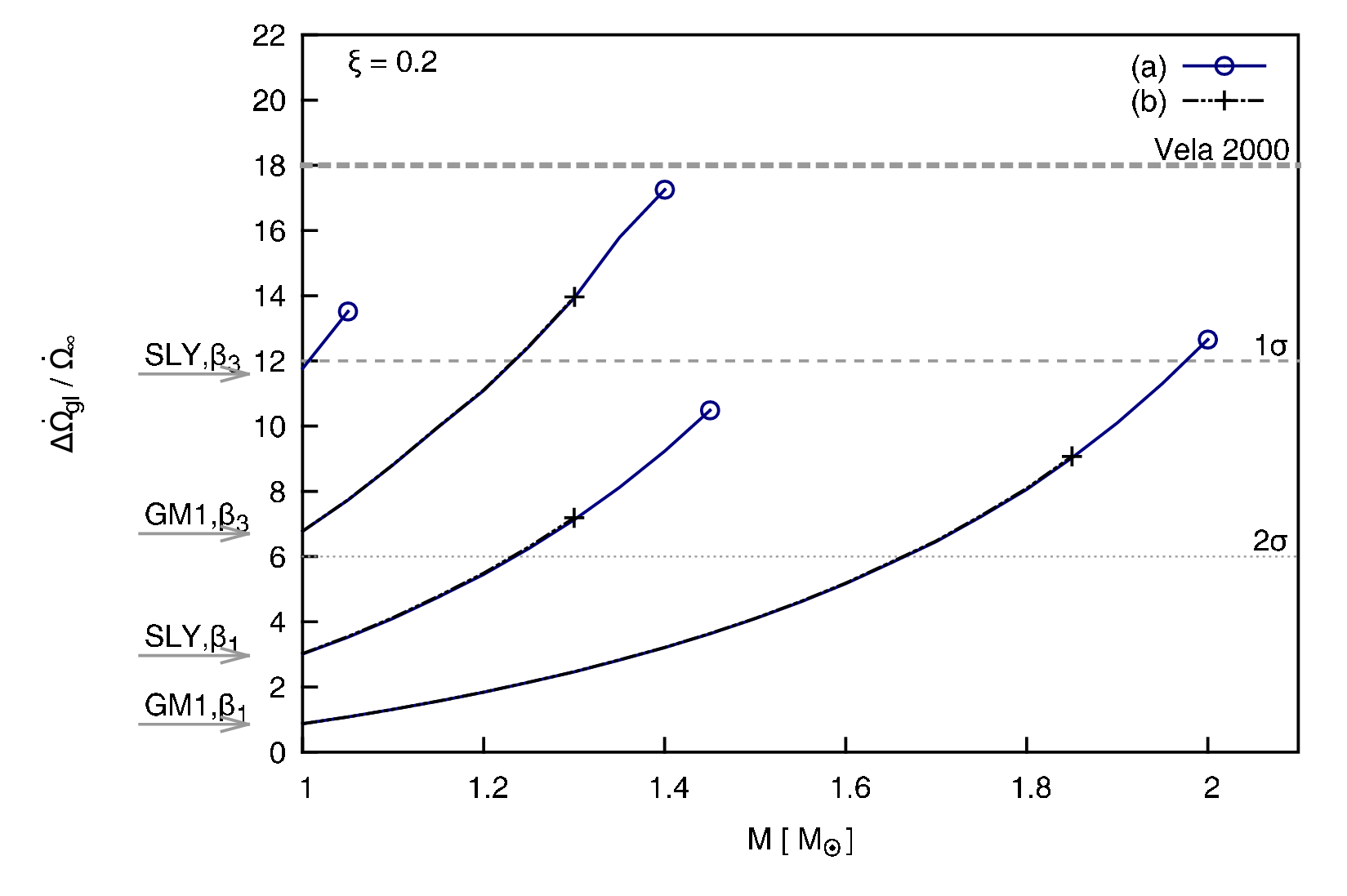}}
\centerline{\includegraphics[scale=0.5]{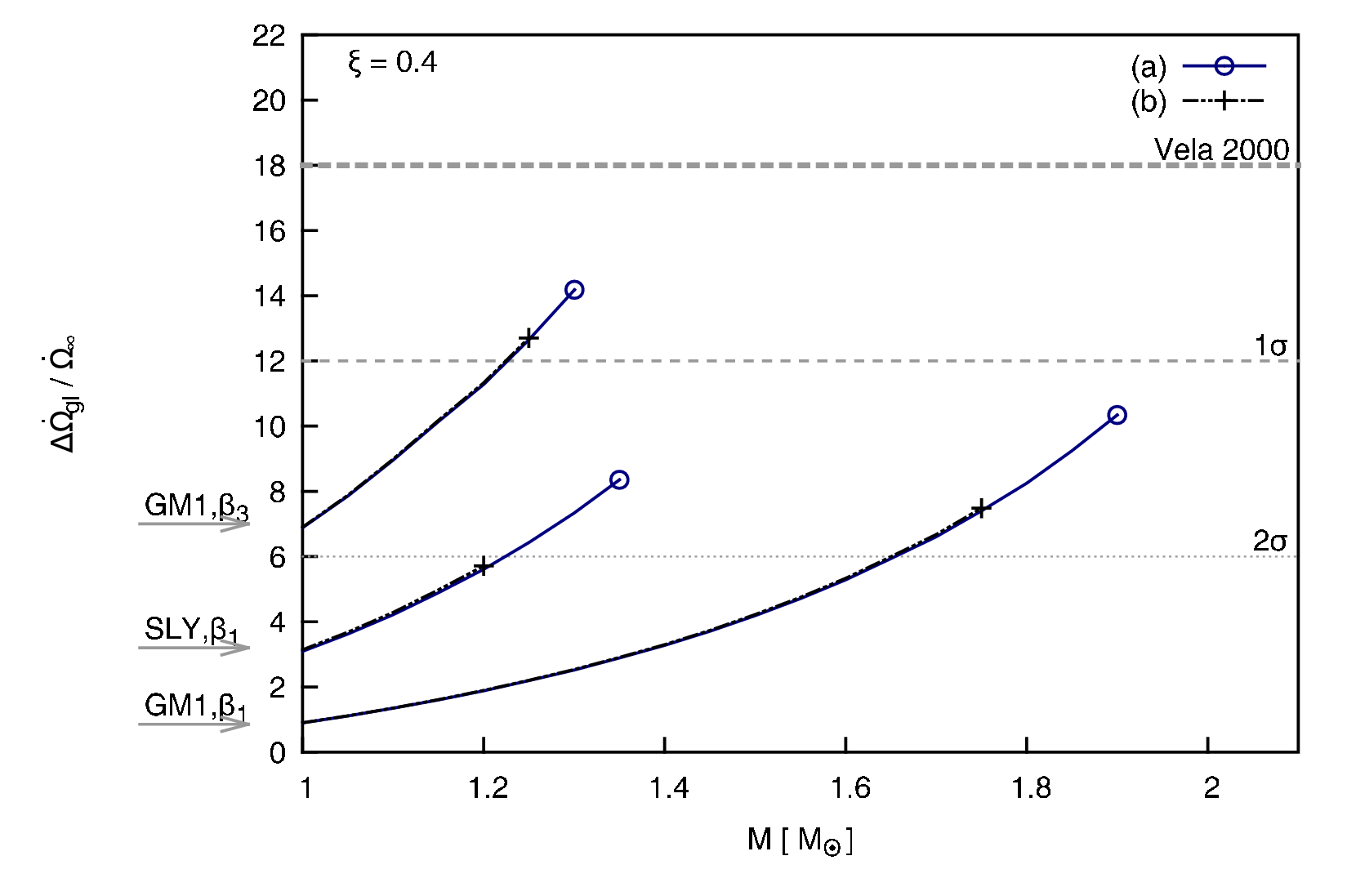}\includegraphics[scale=0.5]{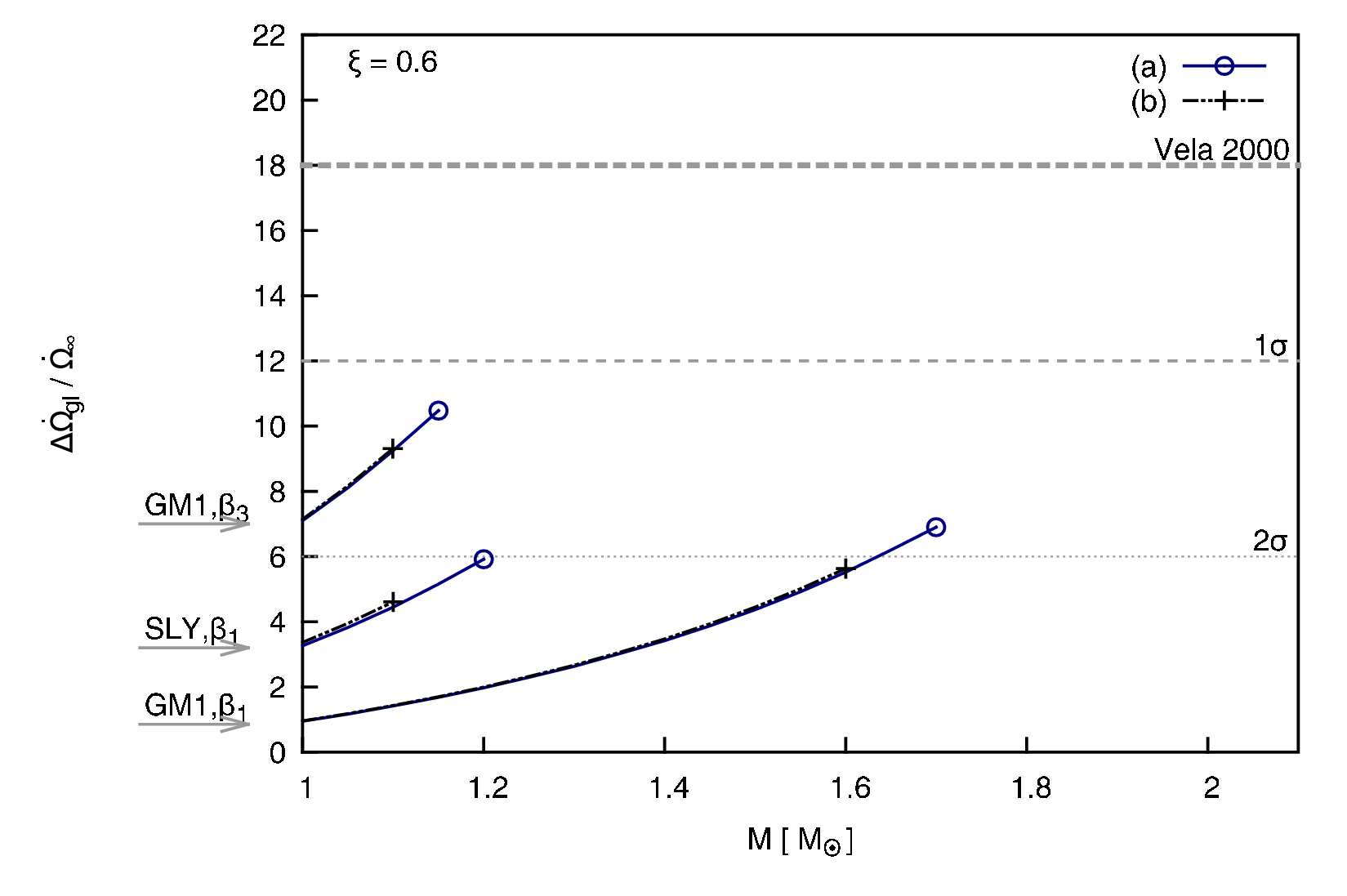}}
\caption{\label{sc1} We plot the value of the fractional step in frequency derivative $\Delta\dot{\Omega}/\dot{\Omega}$ for varying values of the fraction of pinned vorticity in the core, $\xi$ for SLy and GM1. The parameter $\beta$ encodes in-medium polarization effects, as described in the text. For $x_p$ we use the results of \citet{Zuo}, both those obtained with two-body interactions (case a) and with three-body forces (case b). The end of the curves in these two cases corresponds to the point after which we can no longer find a reasonable physical solution. The horizontal line represents the measured value for the Vela 2000 glitch, and the thin lines are respectively 1$\sigma$ and 2$\sigma$ deviations. It is clear that in general both EOSs and models for the proton fraction are compatible with free vorticity in the core. As we increase the pinned fraction however it becomes increasingly more difficult to fit the data, and for $\xi>0.5$ no solution can be found at the 1$\sigma$ level.}
\end{figure*} 

\section{Results}

In order to compare our results with observations we consider the case of the Vela pulsar. The Vela (PSR B0833-45 or PSR J0835-4510) has a spin frequency $\nu\approx 11.19$ Hz and spin-down rate $\dot{\nu}\approx -1.55 \times 10^{-11}$ Hz s$^{-1}$. Giant glitches are observed roughly every thousand days and have relative frequency jumps of the order $\Delta \Omega/\Omega\approx 10^{-6}$. The spin-up is instantaneous to the accuracy of the data, with upper limits of 40s for the rise time obtained from the 2000 glitch \citep{Dod}  and of 30 s for the 2004 glitch, although this limit was less significant \citep{dodson2}). The glitch is usually fitted to a model consisting of permanent steps in the frequency and frequency derivative and a series of transient terms. It is well known that to fit the data at least three are required, with decay timescales that range from months to hours \citep{Flanagan}. Recent observations of the 2000 and 2004 glitch have shown that an additional term is required on short timescales, with a decay time of approximately a minute. Given that the Vela 2000 glitch provides the most robust observational results, we shall compare the expression in (\ref{stepdot}) to the step in frequency derivative associated with the short timescale (1 minute) after the Vela 2000 glitch, which we assume is a reasonable approximation to the instantaneous post-glitch step in the spin down rate. The parameter $Y_{gl}$ is obtained from (\ref{step1}) by fitting to the Vela 2000 glitch size of $\Delta\Omega/\Omega=2.2 \times 10^{-6}$ \citep{Dod}.

In figure (\ref{sc1}) we show the results for varying values of $\xi$, for both SLy and GM1. The parameter $\beta$ encodes the reduction of the pairing gap due to polarization effects in the neutron medium. Recent calculations suggest that polarization reduces the gap and shifts the maximum to lower densities \citep{Gando}, an effect which in our setting corresponds to the value $\beta\approx 3$, while $\beta=1$ corresponds to a bare particle approximation. The horizontal lines show the region that is allowed by the measurements of the step in frequency derivative of the Vela 2000 glitch. It is obvious that most equations of state and proton fractions can match this value if all vorticity in the core is free, as was also found by \citet{Stefano2} and \citet{Hask11}, although we note that for the more realistic case of $\beta=3$ and three body cases included in the calculation of $x_p$, the stiffer equation of state is clearly favored. We now compare this to the case in which part of the core is in a type II superconducting state and part of the vorticity is pinned. As we can see from figure (\ref{sc1}), as the parameter $\xi$ increases it becomes increasingly harder to fit the observed values of $\Delta\dot{\Omega}$ and in most cases this is only possible for a very restricted interval of masses. In general one cannot obtain a physically reasonable fit if more than half of the vorticity in the core is pinned. This points to the conclusion that the vortex/flux tube interaction is weaker than previously assumed and that most of the vorticity in the core is, in fact, free. This conclusion is compatible with that of \citet{Hask11}, who found that a weak coupling between the superfluid and normal component in the core (as would be the case if most of the vortices in the core are pinned) does not allow to fit the shorter post-glitch relaxation timescales of the Vela.
Note that the conclusions of this Letter and those of \citet{Hask11} are derived in different methods (in this case calculating the exchange of angular momentum in a static model, in the case of \citet{Hask11} by fitting the post-glitch relaxation with a dynamical multifluid model) and are thus independent, save for the use of the pinning forces calculated in \citet{Gpaper,Gtesi}.

\section{Conclusions}

In this Letter we have extended the "snowplow" model of \citet{Piz11} to account for the possibility that part of the vorticity in the core may be pinned due to the interaction between vortices and flux tubes. We fit the step in frequency and in frequency derivative of the Vela 2000 glitch to obtain constraints on the pinned fraction of vortices in the core and in general find that both quantities cannot be fitted for reasonable physical parameters if the pinned fraction is larger than 50\%. Although we do not deal with the microphysical details of the vortex dynamics in the core, our conclusions are quite general. The only quantity that is needed to evaluate the angular momentum that is exchanged during a glitch is, in fact, the number of vortices that are stored close to the peak of the pinning potential in the crust. As long as the excess vorticity of the core can be transferred to the equatorial strong pinning region in-between glitches the details of the vortex motion are not influential. 

The general conclusion is that either most of the core is in a type I superconducting state (and the vortex pinning is negligible \citep{Sedrakian05}), or that the vortex/flux tube interaction is weaker than previously thought. This is the same conclusion that \cite{Hask11} come to after fitting the post-glitch short-term relaxation of Vela glitches with a hydrodynamical model. If such an conclusion is confirmed it would have serious implications also for NS precession \citep{Link03} and for GW emission \citep{magnetic, sam, Ho11, Hask12}.
Note that on a microphysical level it is very likely that the interaction between vortices and flux tubes is weaker than the estimates presented here.  These estimates are upper limits on the strength of the pinning force, as they do not account for the finite rigidity of vortices, which could lead to a reduction of a factor 100-1000 \citep{Link11, Gtesi, Gpaper}, as recent theoretical estimates also confirm \citep{SevesoPin}. Furthermore recent calculations \citep{Babaev} suggest that in the presence of strong entrainment or gapped $\Sigma^{-}$ hyperons in the crust the interaction between flux tubes and vortices will be significantly weaker, and even in the presence of pinning the superfluid may be coupled to the crust on short timescales \citep{TrevAlp}.
It should also be pointed out that if flux tubes are able to move out with the neutron vortices on the interglitch timescale, this could lead to a potential barrier at the crust core interface and a glitch \citep{SedCord}. However recent quantitative estimates by \citet{Glamp11} indicate that in the pinning regime vortices and flux tubes can be considered essentially immobile on the inter-glitch timescale, as we assume in the present model.

Finally note that we have assumed straight vortices that cross the core. Superfluid turbulence is, however, a well known phenomenon in laboratory superfluids and is expected to also occur in NSs (see e.g. \citet{Trev07}), especially in the presence of strong pinning \citep{Link11, Link11b}. In this case the vortices will form a turbulent tangle, leading to a weaker interaction between the components and longer coupling timescales in the outer core and crust \citep{Peralta06, Peralta09}, which will however couple the components on long inter-glitch timescales.
Vortex pinning in the core would then be a transitory phenomenon and most of the vorticity would be free on long timescales (see also \citet{Glamp11}).
The transition between a laminar and turbulent flow could however have a strong impact on the glitch mechanism \citep{Peralta06, Peralta07, Peralta09} and the definition of a pinning force per unit length is significantly more complex in a vortex tangle. We plan to tackle this issue in future work.

 \bibliographystyle{apj}  

\end{document}